\documentclass[aps,amsfonts,pra,twocolumn,superscriptaddress,showpacs]{revtex4}
\usepackage{epsfig,amsmath,amssymb,bm,epsf,graphics,psfrag,verbatim}
\def\newB{\lambda_{0}} 
\def\fracstyle{\textstyle}
\def\smhalf{\mbox{\raisebox{0.0pt}{$\fracstyle\frac{1}{2}$}}}

\def\smpi{\mbox{\raisebox{0.0pt}{$\fracstyle\frac{1}{\pi}$}}}
\def\smeighthbeta{\mbox{\raisebox{0.0pt}{$\fracstyle\frac{1}{8T}$}}}
\def\smhbetaB{\mbox{\raisebox{0.0pt}{$\fracstyle\frac{\newB}{2}$}}}
\def\density{\mbox{\raisebox{0.0pt}{$\fracstyle\frac{N}{V_0}$}}}
\def\Gibbs{\Gamma}
\def\GibbsMicro{{\cal G}}
\def\tz{\tilde{z}}
\def\tp{\tilde{p}}
\def\tq{\tilde{q}}
\def\tR{\widetilde{R}}
\def\tu{\tilde{u}}
\def\tG{\widetilde{G}}
\def\teps{\tilde{\varepsilon}}
\def\tpart{\widetilde{\partial}}

\def\cld{\eta^2}    
\def\myQ{Q}         
\def\myP{\wp}       
 \def\delbar{{\mathchar'26\mkern-9mu\delta}}    
\def\phT{^{\phantom{\textrm{T}}}}
\def\dlrho{\theta}
\begin{document}
\title{Elastic heterogeneity of soft random solids}
\author{Xiaoming Mao}
\affiliation{Department of Physics,
University of Illinois at Urbana-Champaign,
1110 West Green Street, Urbana, Illinois 61801
}

\author{Paul M.~Goldbart}
\affiliation{Department of Physics,
University of Illinois at Urbana-Champaign,
1110 West Green Street, Urbana, Illinois 61801
}

\author{Xiangjun Xing}
\affiliation{Department of Physics,
Syracuse University, Syracuse, New York 13244
}

\author{Annette Zippelius}
\affiliation{Institut f\"ur Theoretische Physik,
Universit\"at G\"ottingen,
37077 G\"ottingen, Germany}
  \date{October 15, 2006}

\begin{abstract}
Spatial heterogeneity in the elastic properties of soft random solids is investigated via a two-pronged approach.  First, a nonlocal phenomenological model for the elastic free energy is examined.  This features a quenched random kernel, which induces randomness in the residual stress and Lam\'e coefficients.  Second, a semi-microscopic model network is explored using replica statistical mechanics.  The Goldstone fluctuations of the semi-microscopic model are shown to reproduce the phenomenological model, and via this correspondence the statistical properties of the residual stress and Lam\'e coefficients are inferred.  Correlations involving the residual stress are found to be long-ranged and governed by a universal parameter that also gives the mean shear modulus.
\end{abstract}
\maketitle

\noindent
{\it Introduction\/}---%
As a consequence of randomness incorporated at synthesis, random solids (e.g.~polymer networks, glasses, $\alpha$-Si) are heterogeneous.
For example, the mean positions of the constituent particles exhibit no apparent long-range order, and every particle inhabits a unique spatial environment.
Particularly for {\it soft random solids\/}, such as rubber, in which the particle positions undergo large thermal fluctuations,  heterogeneity also manifests itself via the RMS particle-displacements, which are random and continuously distributed~\cite{REF:castillo94,REF:CGZ-AdvPhy}.

The elasticity of rubber, and especially its softness with respect to shear deformations, have been studied for many years via the classical theory, developed by Kuhn, Flory, Wall, Treloar and others~\cite{REF:Treloar} and based on a microscopic picture of Gaussian polymer chains.
While the classical theory has proven highly successful, it is a homogeneous theory, preserving no information about the random structure of these essentially heterogeneous solids.
Thus, it is incapable of describing consequences of the heterogeneity, such as random spatial variations in the local elastic parameters and residual stress, or non-affine deformations that occur in reponse to applied stresses~\cite{REF:nonaffine}.

The purpose of the present Letter is to present a theoretical development that goes beyond the classical theory of rubber elasticity by accounting for the heterogeneity.
What emerges is an elasticity theory featuring spatially fluctuating Lam\'e coefficients and residual stresses, together with a statistical characterization in terms of their mean values and spatial self- and cross-correlations.
In particular, we find that not only is the stress-stress correlation long ranged---behavior that can be argued for on general grounds---but so are the cross-correlation between the residual stress and, e.g., the shear modulus.
By contrast, we find the self-correlation of the shear modulus to be short-ranged.
Furthermore, we find that the long-ranged correlations and the average shear modulus are governed by a common universal parameter that is independent of microscopic details.

To obtain our statistical characterization of soft random solids we take the following route.
First, we examine a nonlocal phenomenological model of a random elastic medium, which we subsequently derive from a semi-microscopic model. We then determine the state to which it relaxes when randomness is present, and re-expand the elastic free energy around this new equilibrium reference state~\cite{REF:DiDonna2005-use}.
This relaxed state is, however, still randomly stressed~\cite{REF:Alexander}; nevertheless, the stress in the relaxed state---the so-called \emph{residual stress}---satisfies the mechanical equilibrium condition $\partial _j \sigma_{jk}(x)=0$.
In its local limit, the proposed phenomenological model reproduces a version of Lagrangian elasticity theory that features random Lam\'e coefficients and residual stresses.
Second, we consider the statistical mechanics of a certain semi-microscopic model of a random-solid-forming system---the randomly linked particle model (RLPM)~\cite{REF:GOE-BPM}.
Via the replica method, applied to the RLPM to deal with its structural randomness, followed by an analysis of Goldstone fluctuations of the random solid state of the RLPM, we arrive at {\it precisely\/} the aforementioned nonlocal phenomenological model, except that the latter has been appropriately disorder-averaged using replicas.
By comparing these two disorder-averaged models, we characterize the elastic heterogeneity of soft random solids.

\noindent
{\it Phenomenological model\/}---%
We begin by examining the following nonlocal model for the elastic free energy $\Gibbs$ of soft random solids, which describes \lq\lq mass points\rq\rq\ (i.e.~coarse-grained volume-elements) that interact with one another through random harmonic attractions:
\begin{eqnarray}
\Gibbs\!\!&=&\!\!
\smhalf \!\int\! dz_{1}\,dz_{2}\,
G(z_{1},z_{2})
\Big\lbrace
\big\vert R(z_{1})-R(z_{2})\big\vert^{2}\!\!-
\big\vert z_{1}-z_{2}\big\vert^{2}
\Big\rbrace
\nonumber \\
&&\qquad\qquad\qquad+\,\smhalf\,\newB\!\int \!dz\,
\big(\left\vert{\partial R(z)}\right\vert-1\big)^{2},
\label{EQ:phenom_model}
\end{eqnarray}
i.e., a functional of the deformation of the system $R(z)$, which specifies the $D$-dimensional position vector to which the mass point at position $z$ is displaced.
$\vert{\partial R(z)}\vert$
denotes the determinant of the deformation gradient tensor
$\partial R_{i}/\partial z_{j}$
and, correspondingly, the parameter $\newB$, which we take to be large, heavily penalizes density variations.
It results from a competition between (i)~repulsions (either direct or mediated via solvent, e.g., excluded-volume) and (ii)~intermolecular attractions and external pressure.
The nonlocal kernel $G(z_{1},z_{2})$ describes link-induced harmonic attractions between mass points, originating in the entropy of the molecular chains of the heterogeneous network, and are modeled as a \lq\lq zero-rest-length\rq\rq\ spring of random spring coefficient.  $G(z_{1},z_{2})$ is taken to be a quenched random function of the two positions, $z_{1}$ and $z_{2}$, symmetric under $z_{1}\leftrightarrow z_{2}$.  This model, which we shall shortly see emerging from the RLPM, is in the spirit of the classical theory of rubber elasticity.

The free energy $\Gibbs$ is not stable at the state $R(z)=z$ for two reasons:
first, the attraction $G$ causes a small, spatially uniform contraction [the fractional volume change being ${\cal O}(1/\newB)$];
second, the randomness of $G$ additionally destabilizes this contracted state, causing the adoption of a randomly deformed equilibrium state.
We denote this relaxation as $z\to\tz=\zeta z +v(z)$, in which $\zeta$ describes the uniform contraction, and $v(z)$ describes the random local deformation.
This process can be understood in the setting of the preparation of a sample of rubber via instantaneous crosslinking: crosslinking not only drives the liquid-to-random-solid transition but also generates a uniform inward pressure, as well as introducing random stresses.
As a result, immediately after crosslinking the state is not stable, but relaxes into a new equilibrium state determined by the particular realization of randomness created by the crosslinking.
Free energy stationarity applied to Eq.~(\ref{EQ:phenom_model}) shows that at large $\newB$ the contraction
$\zeta\approx 1-({\rho}/{D\newB})$,
where
$\rho\equiv
\frac{1}{D}
\int dz_2\,(z_1-z_2)^2\,G_0(z_1-z_2)$,
and $G_{0}(z_{1}-z_{2})\equiv [G(z_{1},z_{2})]$ is the disorder average (denoted by $\lbrack\cdots\rbrack$) of $G$, and we have assumed that this disorder averaged $G$ is translationally and rotationally invariant.
As we shall see below, $\rho$ is actually the mean shear modulus.

The random relaxation $v(z)$ is, in general, difficult to determine, owing to the large nonlinearity from the $\newB$ term.  However, by assuming that $G_{1}\equiv G-G_0$ (the random part of $G$) is {\it small\/}, and seeking $v(z)$ to first order, free energy minimization \(\delta\Gibbs/\delta v(z)=0\) enforces that the Fourier transform
$v(p)\equiv \int dz\,\exp(-ipz)\,v(z)$ is given by
\begin{eqnarray}
v(p)&=&\!
\frac{{\cal P}^{\text{T}}(p) \cdot f(p) }{2\big(G_{0}(0)\!-\!G_{0}(p)\big)}+
\frac{{\cal P}^{\text{L}}(p) \cdot f(p)}{\newB p^{2}\!+2\big(G_{0}(0)\!-\!G_{0}(p)\big)},
\qquad
\label{EQ:v-solved}\\
\noalign{\smallskip}\nonumber
\end{eqnarray}
where $f(z)\equiv -2\zeta \int dz_{1}\,G_{1}(z,z_{1})\,(z-z_{1})$,
and ${\cal P}^{\text{L}}_{jk}(p)\equiv p_{j}\phT\,p_{k}\phT/p^{2}$
and ${\cal P}^{\text{T}}_{jk}(p)\equiv \delta_{jk}\phT-{\cal P}^{\text{T}}_{jk}(p)$
are, respectively, transverse and longitudinal projector, and $j$, $k$, etc.~are Cartesian indices.

Next, we transform to a new reference state, which we take to be the relaxed state, via the coordinate transformation $z \to \tz (z)= \zeta z+v(z)$ [with inverse $z(\tz)$].
Correspondingly, we obtain the relaxed-state kernel
$\tG\big(\tz_{1},\tz_{2}\big)\equiv G\big(z(\tz_{1}),z(\tz_{2})\big)$.
We then re-expand the elastic free energy as a functional of the deformation $\tR(\tz)=\tz+\tu(\tz)$ around the new reference state, keeping terms to quadratic order in the nonlinear Lagrangian strain tensor
$\teps_{jk}(\tz)
\equiv
\frac{1}{2}
\big(
(\partial\tR_{l}/\partial\tz_{j})
(\partial\tR_{l}/\partial\tz_{k})
-\delta_{jk}
\big)$
and to sub-leading order in $1/\newB$.
By considering the local limit of the resulting free energy via a gradient expansion, and dropping an additive constant, we arrive at
\begin{equation}
\Gibbs\!=\!\!
\int\!\!d\tz
\Big\lbrace
\text{Tr}\,\sigma(\tz)\cdot\teps(\tz)
\!+
\mu(\tz)\text{Tr}\,\teps(\tz)^2
\!+
{\mbox{\raisebox{0.0pt}{$\fracstyle\frac{\lambda(\tz)}{2}$}}}
\big(\text{Tr}\,\teps(\tz)\big)\!^{2}\!
\Big\rbrace.\!
\label{EQ:local_hetero}
\end{equation}
This free energy features three random elements: the heterogeneous Lam\'e coefficients $\mu(\tz)$ and $\lambda(\tz)$
and the residual stress $\sigma_{jk}(\tz)$; respectively, they are given by
\begin{subequations}
\label{EQS:mu-lambda}
\begin{eqnarray}
&&\mu(\tp)
\equiv
\rho
           \,\delbar(\tp)
-i\zeta^{-1} \tp^{-2} (\tp \cdot f(\tp)),
\\
\noalign{\smallskip}
&&\lambda(\tp)
\equiv
\lambda _0
           \,\delbar(\tp)
 +2\big(i\zeta^{-1}\tp^{-2} (\tp \cdot f(\tp))
 -\rho
           \,\delbar(\tp)
 \big),
\\
\noalign{\smallskip}
&&\sigma_{jk}(\tp)
\equiv
-\frac{\partial^2}{\partial \tq_j\,\partial \tq_k}
 \Big\vert_{\tq =0} G_1(\tp-\tq,\tq)
+ i\delta_{jk}\frac{\tp \cdot f(\tp)}{\zeta {\tp}^{2}}
\nonumber\\
\noalign{\smallskip}
&&\qquad+
 i\zeta^{-1}\tp^{-2}
 \big(\tp_j\phT \mathcal{P}^{\textrm{T}}_{kl}(\tp)+
      \tp_k\phT \mathcal{P}^{\textrm{T}}_{jl}(\tp)\big)
   f_l\phT(\tp),
\label{EQ:comp-stress}
\end{eqnarray}
\end{subequations}
where $\delbar(\tp)\!\equiv(\!2\pi)^D \delta(\tp)$.
It can be shown that $\sigma$ obeys the equilibrium condition $\tpart_{j}\sigma_{jk}(\tz)=0$.\thinspace\
(Strictly speaking, the residual stress $\sigma$, defined here, is the leading-order term of the true stress, in the sense of a gradient expansion, owing to the non-locality of our model.  As a result, the equilibrium condition only holds to the corresponding order in the gradient expansion.)

To determine the statistics of $G$, we first treat the statistical mechanics of elastic fluctuations governed by Eq.~(\ref{EQ:phenom_model}) using replicas. [We explain why this procedure is appropriate, below, before Eq.~(\ref{EQ:identify}).]\thinspace\ Thus, we arrive at the disorder-averaged replicated partition function
\begin{eqnarray}
\big[Z^{n}\big]&=&
\int\prod\nolimits_{\alpha=1}^{n}{\cal D}R^{\alpha}\,\exp{(-\Gibbs_n/T)},
\end{eqnarray}
where we have set $k_{\rm B}=1$ and
$\alpha$ labels replicas;
$[Z^{n}]$ is related to the disorder averaged Gibbs free energy via
$[\Gibbs]=-T\lim_{n\to 0}\frac{\partial}{\partial n}\ln [Z^n]$
~\cite{REF:BY-1986}.
The resulting effective pure free energy is
\begin{eqnarray}
&&
\!\!\!\!\!\!\!
\Gibbs_n
\!=
\smhbetaB\!\!
\int\nolimits_{z}\sum_{\alpha=1}^{n}
\big(\left\vert{\partial R^{\alpha}}\right\vert-1\big)^{2}
\!\!
+\smhalf\!\!
\int\nolimits_{\{z_{i}\}}\!\!
\!\! [G(z_{1},z_{2})]\,
\Psi_{R}(z_1,z_2)
\nonumber\\
&&
\!\!\!\!\!\!\!
-\smeighthbeta\!\!
\int\nolimits_{\{z_{i}\}}\!\!
\!\!\!\!
[G(z_{1},z_{2})G(z_{3},z_{4})]_{\rm c}\,\!
\Psi_{R}(z_1,z_2)
\Psi_{R}(z_{3},z_{4})
\!+\!\cdots \!,
\label{eq:HnPh}
\end{eqnarray}
where the subscripts $\{z_{i}\}$ indicate integration variables, \lq c\rq\ indicates a cumulant (or connected correlator), and
\begin{equation}
\Psi_{R}(z_1,z_2)\equiv\sum _{\alpha=1}^{n} \big\lbrace
\vert R^{\alpha}(z_1)-R^{\alpha}(z_2)\vert ^2 - \vert z_1-z_2 \vert ^2\big\rbrace.
\end{equation}
The dots represent terms arising from higher-order cumulants, which characterize the non-Gaussian nature of the distribution of $G$.

\noindent
{\it Semi-microscopic model\/}---%
Our next goal is to determine the cumulants via the statistical-mechanical analysis of a semi-microscopic model.
Thus, we consider the randomly linked particle model (RLPM)~\cite{REF:GOE-BPM}, which consists of $N$ particles having coordinates $\{c_j\}_{j=1}^{N}$ interacting via an excluded-volume term, all in a fluctuating volume $V$, the mean value of which is controlled by a pressure $p$.
In addition, permanent soft links, modeled as \lq\lq zero-rest-length\rq\rq\ springs, are introduced at random, with a separation-dependent probability, between nearby particles in a liquid-state configuration, so that the number of links is quasi-Poisson-distributed and the correlations among the links are consistent with the correlations of the liquid state.
This is a version of the Deam-Edwards distribution~\cite{REF:DE-1975}, and is at least appropriate for instantaneous schemes for synthesizing soft random solids.

For a given realization of the randomness, the Hamiltonian of the RLPM is given by
\begin{eqnarray}
H_{\rm RLPM}=
\frac{\nu ^2}{2}
\sum\nolimits_{i,j=1}^{N}
\delta (c_i-c_j)+
\sum\nolimits_{e=1}^{M}
\frac{\vert c_{i_e}-c_{j_e}\vert ^2}{2a^2}.
\end{eqnarray}
The first term describes excluded-volume repulsion ($\nu^2$ is taken to be large and, thus, density variations are suppressed); the second term describes attractions associated with the $M$ randomly-chosen pairs $\{c_{i_e},c_{j_e}\}_{e=1}^{M}$ of particles that have been softly linked to one another.

The particles of the RLPM can be identified with small molecules or coarse-grained polymers, and the soft links, e.g.,  with Gaussian molecular chains.
In making this coarse-graining one is assuming that microscopic details (e.g., the precise positions of the crosslinks on a polymer, the internal conformational degrees of freedom of the polymers, and the effects of entanglement) do not play significant roles for the long wave-length physics.
In part, these assumptions are justified by studying more detailed models, in which the conformational degrees of freedom of the polymers are retained~\cite{REF:CGZ-AdvPhy}.
Thus, the RLPM can be regarded as a caricature of vulcanized rubber or as a model of chemical gels or other soft random solids.
It is a model very much in the spirit of lattice percolation, except that it naturally allows for particle motion as well as connectivity, and is therefore suitable for the study of continuum elasticity and other issues associated with the (thermal or deformational) motion of particles.

To analyze the RLPM we use replica field theory, which allows us to average over the quenched randomness, and, accordingly, we introduce the following order parameter, suitable for describing the soft random solid state:
\begin{eqnarray}
\Omega (\hat{x})=\frac{1}{N}\!\sum_{i=1}^{N} [
\langle \delta(x^0-c_i) \rangle
\langle \delta(x^1-c_i) \rangle \ldots
\langle \delta(x^n-c_i) \rangle
].
\nonumber
\end{eqnarray}
Here, $\hat{x}=(x^0,x^1,\ldots,x^n)$ denotes the $(1+n)$-fold-replicated $D$-vector $x$, and $\langle \cdots \rangle$ denotes a thermal average for a given realization of randomness (i.e.~a given configuration of the links)~\cite{REF:CGZ-AdvPhy}.
This order parameter is capable of detecting and characterizing the random solid state.
The additional (i.e.~0$^{\rm th}$) replica arises from the preparation ensemble associated with the Deam-Edwards distribution of the quenched randomness; this replica has a fixed volume $V_{0}$.
We fix the pressure in the {\it measurement\/} ensemble to equal the mean pressure in the {\it preparation\/} ensemble, i.e., $p=\frac{\nu^2 N^2}{2V_0^2}+\frac{NT}{V_0}$.  This leads to the following effective pure $\Omega$-dependent free energy:
\begin{eqnarray}
    &&\!\!\!\!
    \GibbsMicro_{1+n}=\frac{NT\cld}{2V^n \Delta_0} \!\! \sum_{\hat{p} \! \in \textrm{HRS}}
    \Delta (\hat{p}) \Omega (\hat{p}) \Omega(-\hat{p})+\frac{n{\tilde\nu}^{2}N^2}{2V}+npV \,\,\,
    \nonumber
    \\
    &&\qquad\!\!\!\!
    -NT\ln\int d\hat{c}
    \exp\Big(
    \frac{\cld}{V^n} \!\! \sum_{\hat{p} \in \textrm{HRS}}
    \Delta (\hat{p}) \Omega (\hat{p})  e^{i\hat{p}\cdot \hat{c}}
    \Big).
\end{eqnarray}
Here, $\Omega (\hat{p})$ and $\Delta (p)$ are, respectively, the Fourier transforms of $\Omega (\hat{x})$ and $\Delta(x)$ ($\equiv\exp(-{x^2}/{2a^2})$), and $\Delta (\hat{p})\equiv\prod_{\alpha=0}^{n}\Delta (p^{\alpha})$ and $\Delta_0=\Delta(p)\vert _{p=0}$.
Then, the disorder-averaged Gibbs free energy is given by
\begin{equation}
[\GibbsMicro]=-T\lim_{n\to 0}
\frac{\partial}{\partial n}\ln\int{\cal D}\Omega\,\exp\left(-\GibbsMicro_{1+n}/T\right).
\end{equation}
The parameter $\cld$ controls the {\it density\/} of links, with the average number of links to any single particle being $\cld$.
The restriction $\hat{p}\in\textrm{HRS}$ serves to exclude macroscopic density fluctuations, i.e., to account for the effects of the strong excluded-volume interactions~\cite{REF:CGZ-AdvPhy}.
At the level of mean-field theory, the transition from the liquid to the random solid state occurs at $\cld=1$, as can readily be seen via an expansion of $\GibbsMicro_{1+n}$ in powers of $\Omega$.
For larger $\cld$, the equilibrium value of the order parameter is
\begin{eqnarray}\label{EQ:SPOP}
\Omega(\hat{x})=
\!\myQ\!\!\int\!\frac{dz}{V_0}\!
\int\!\!d\tau\,\myP(\tau)\,
e^{-\frac{\tau}{2}\{\vert{x^{0}-z}\vert^{2}
-\sum_{\alpha=1}^{n}\vert{x^{\alpha}-\zeta z}\vert^{2}\}},
\end{eqnarray}
where $\hat{z}\equiv(z,\zeta z,\ldots,\zeta z)$ represents the random mean positions about which replicated particles
undergo thermal fluctuations, $\zeta=(V/V_0)^{1/d}$ is the contraction ratio,
and $\myQ$ and $\myP(\tau)$ characterize the state via, respectively, the fraction of localized particles and the distribution of (inverse square) localization lengths.
Again within mean-field theory, $\myQ$, $\myP$ and $\zeta$ are determined by self-consistency conditions that follow from making $\GibbsMicro_{1+n}$ stationary; in particular, $\myQ$ obeys
$1-\myQ=\exp (-\cld\myQ)$~\cite{REF:castillo94,REF:CGZ-AdvPhy}.

With the approach to the RLPM in place, we now describe how the equilibrium value of the order parameter is modified when the system undergoes an elastic deformation.
In this replica theory, such deformations amount to a replica version of Goldstone excitations. In view of the pattern of spatial symmetry breaking, they are specified by $n$ (not $n+1$) independent long wavelength deformation fields $\{R^{1}(z),R^{2}(z),\ldots,R^{n}(z)\}$, each depending on a single replica of space  $z$~\cite{REF:MGZ-2004,REF:newgold}, with the constraint ${\rm det}\,(\partial R^{\alpha}/\partial (\zeta z))=1$. This constraint corresponds to the low-energy excitations of the model: pure shear deformations from the contracted measurement state.
Hence, we arrive at the appropriately deformed order parameter:
\begin{eqnarray}
\myQ\int \! \frac{dz}{V_0}
\int \! d\tau \, \myP(\tau)
\exp\big({-{\tau\vert\hat{x}-\widehat{R}(z)\vert^2}/{2}}\big),
\end{eqnarray}
where $\widehat{R}(z)\!\equiv\!\big(z,R^1(z),\ldots,R^n(z)\big)$ are the random mean positions of the particles in the preparation ($\alpha\!=\!0$) and (deformed) measurement ($\alpha\!=\!1,\ldots,n$) replicas, and
$\hat{x}^2$ denotes $\sum_{\alpha=0}^{n}\vert x^\alpha \vert ^2$.
The link density and excluded-volume parameters will be taken to obey: ${\tilde\nu}^{2}N/TV_{0}\gg\cld\gg 1$.  We take $\cld\gg 1$ (and hence $\myQ\approx 1$) because we shall be concerned with the well-linked regime, in which only a very small fraction of particles are not part of the infinite cluster.  As we shall see, the condition ${\tilde\nu}^{2}N/TV_{0}\gg 1$ indicates that the medium is near-incompressible, and ${\tilde\nu}^{2}N/TV_{0}\gg\cld$ holds because we are concerned with {\it soft\/} random solids, for which the bulk modulus greatly exceeds the shear modulus.

To determine the free energy
of the Goldstone deformation we insert the deformed order parameter into $\GibbsMicro_{1+n}$, thus obtaining (up to an irrelevant additive constant)
\begin{eqnarray}
&&\!\!\!\!\!\!
\GibbsMicro_{1+n}[\widehat{R}]=
\frac{\nu^{2}N^{2}}{2V_{0}}
\left(\frac{V}{V_{0}}-1\! \right)^{2}
\!\!\!
+\smhalf\!\int_{\{z_i\}}
\!\!\!\!\!
K_1(z_1,z_2)\,\Psi_{R}(z_1,z_2)
\nonumber
\\
&&\!\!\!\!\!
-{\scriptstyle\frac{\scriptstyle 1}{\scriptstyle 8T}}\!\!
\int_{\{z_i\}}\!\!\!\!\!
K_2(z_{1},z_{2},z_{3},z_{4})
\Psi_{R}(z_{1},z_{2})\,
\Psi_{R}(z_{3},z_{4})\!+\!\cdots.\!\!
\label{EQ:BPM-goldstone}
\end{eqnarray}
The kernels $ K_1(z_1,z_2) $ and $K_2(z_{1},z_{2},z_{3},z_{4})$ are given by rather lengthy formul\ae\ in terms of $\cld$, $\myQ$ and $\myP$, but are, in essence, bell-shaped functions of the separations of the variables that fall off on the scale of the typical localization length; we shall report on them elsewhere.

\noindent
{\it Results\/}---%
Observe that Eqs.~(\ref{eq:HnPh}) and (\ref{EQ:BPM-goldstone}) are equivalent, in the sense that they are both capable of describing the free energy (disorder-averaged via replicas) of spatially varying deformations and homogeneous volume variations in soft random solids, about an unrelaxed reference state $R(z)=z$.  This equivalence makes the RLPM a sensible candidate for addressing the question posed above about the statistics of $G_{1}$.  By comparing the these two equations we arrive at the correspondence
\begin{subequations}
\label{EQ:identify}
\begin{eqnarray}
[G(z_1,z_2)]_{\phantom{\rm c}}&=&
K_1(z_1,z_2),\\
\lbrack G(z_{1},z_{2})\,G(z_{3},z_{4})\rbrack _{\rm c}
&=&
K_2(z_{1},z_{2},z_{3},z_{4}),\\
\newB&=&\nu^{2}N^{2}/V_{0}^{2},
\end{eqnarray}
\end{subequations}
Higher-order cumulants of $G$ may be determined via the expansion of $\GibbsMicro[\widehat{R}]$ to higher order in 
$\Psi_{R}$, as we shall undertake elsewhere.

\def\myshift{\,\,}
\begin{table}[h]
\begin{tabular}{|r|c c c|}
\toprule
    &$\myshift\sigma_{kl}(p^{\prime})\myshift$
    &$\myshift\mu(p^{\prime})\myshift$&
     $\myshift\lambda(p^{\prime})\myshift$
\\
\hline
     $\myshift            \sigma_{ij}(p)\myshift$
    &$\myshift           2\dlrho A_{ijkl}\myshift$
    &$\myshift          -4\dlrho\mathcal{P}^{\rm T}_{ij}(p)\myshift$
    &$\myshift\phantom{-}8\dlrho\mathcal{P}^{\rm T}_{ij}(p)\myshift$
\\
     $\myshift            \mu(p)\myshift$
    &$\myshift          -4\dlrho\mathcal{P}^{\rm T}_{kl}(p)\myshift$
    &$\myshift\phantom{-2} \gamma\myshift$
    &$\myshift          -2 \gamma\myshift$
\\
     $\myshift            \lambda(p)\myshift$
    &$\myshift\phantom{-}8\dlrho\mathcal{P}^{\rm T}_{kl}(p)\myshift$
    &$\myshift          -2 \gamma\myshift$
    &$\myshift\phantom{-}4 \gamma\myshift\myshift$
\\
\botrule
\end{tabular}
\caption{
Long-wavelength variances and covariances of the elastic properties of soft random solids in the relaxed state.
The entry in row R and column C, when multiplied by
$(N/V_{0})T^2 \delbar(p+p^{\prime})$,
yields the connected disorder correlator
$[{\rm R}(p)\,{\rm C}(p^{\prime})]_{\rm c}\equiv
[{\rm R}(p)\,{\rm C}(p^{\prime})]-
[{\rm R}(p)]\,[{\rm C}(p^{\prime})]$.
}
\label{TAB:variances}
\end{table}

Now that we have ascertained information about the statistics of $G$, we use it to address the statistics of the heterogeneous Lam\'e coefficients $\mu$ and $\lambda$ and the residual stress $\sigma$ of the relaxed state, which we do via Eqs.~(\ref{EQ:v-solved}) and (\ref{EQS:mu-lambda}).
This lead to the following results for the disorder-averaged elastic parameters:
\begin{equation}
\big(\,\lbrack\sigma _{jk}(\tz)\rbrack,\,
\lbrack\mu (\tz)\rbrack,\,
\lbrack\lambda(\tz)\rbrack\,\big)=
\big(0\,,\rho \,,\nu^{2}N^{2}/V_{0}^{2} \, \big),
\end{equation}%
where the average shear modulus $\rho$ is given by~\cite{REF:Ulrich-2006}
\begin{equation}
\rho=\dlrho TN/V_{0},
\quad
\dlrho \equiv
-\smhalf\, {\cld  \myQ^2 }
+e^{-\cld  \myQ }
+\cld  \myQ -1.
\end{equation}
$\dlrho\sim(\cld-1)^{3}$ near the transition, and $\dlrho\sim\cld$ for $\cld\gg 1$.
The variances and covariances among the Fourier transforms of $\mu$, $\lambda$ and $\sigma$ at long wavelengths are given in Table~\ref{TAB:variances}.
In these formul\ae, $\gamma$ is a scalar determined by $\cld$ and $\myP$; the tensor $A$ is a $p$-dependent structure that vanishes on contraction with $p$.

By transforming these variances and covariances back to real space, one can obtain their leading large-distance behavior.  It is evident that the entities {\it not\/} involving the stress $\sigma(r)$
(i.e.~$\lbrack\mu(0)    \mu(r)    \rbrack_{\rm c}$,
      $\lbrack\lambda(0)\lambda(r)\rbrack_{\rm c}$ and
      $\lbrack\mu(0)    \lambda(r)\rbrack_{\rm c}$)
are short-ranged in real space: more precisely, they are proportional to $\delta_{\rm s}(r)$
(i.e.~to the Dirac delta function smoothed on the scale of the short-distance cutoff, which should be taken to be the typical scale for localization length, in order to validate the Goldstone-fluctuation framework).
By contrast, those entities involving the stress have more interesting behavior: in three dimensions and at large-scales we find that
\begin{subequations}
\begin{eqnarray}
\lbrack\sigma _{ij}(0)\,\sigma_{kl}(r)\rbrack_{\rm c}
    \!\!&=&\!\!\! \phantom{-}\smpi\,\dlrho T^{2}\density B _{ijkl}/r^3,
    \\
    \lbrack\sigma _{ij}(0)\,\mu(r)\rbrack_{{\rm c}}
    \!\!&=&\!\!\!
    -\mbox{\raisebox{0.0pt}{$\fracstyle\frac{2}{\pi}$}}\,\dlrho T^{2}\density
    \left(\mathcal{P}_{ij}^{\rm L}(r)-\smhalf\mathcal{P}_{ij}^{\rm T}(r)\right)\!/r^3,
    \quad\quad
    \\
    \lbrack\sigma _{ij}(0)\,\lambda(r)\rbrack_{{\rm c}}
    \!\!&=&\!\!\!
    \phantom{-}
    \mbox{\raisebox{0.0pt}{$\fracstyle\frac{4}{\pi}$}}\,\dlrho T^{2}\density
    \left(\mathcal{P}_{ij}^{\rm L}(r)-
    \smhalf \mathcal{P}_{ij}^{\rm T}(r)\right)\!/r^3,
    \quad
\end{eqnarray}
\end{subequations}
where the tensor $B_{ijkl}$ is a complicated structure comprising terms built from projection operators of the vector $r$, together with various index combinations.  It is perhaps unexpected that the correlation between the residual stress and shear modulus is a long-ranged quantity.

By starting with a microscopic model, we have constructed a random nonlocal elasticity theory, together with a statistical characterization of the parameters that feature in this theory.
So far, we have focused on this statistical characterization in the {\it local\/} limit of the theory, Eq.~(\ref{EQ:local_hetero}), and have found---inter alia---that a universal parameter $\dlrho$ controls all long-range correlations.  This parameter also governs the large-distance statistics of non-affine deformations that occur in response to applied stresses~\cite{REF:DiDonna2005-ref}.
The statistical content and implications of the theory can also be explored {\it beyond\/} the local limit.  For example, the complete statistics of the nonlocal kernel $G$ are amenable to the present formalism, and will allow the extension of the statistics of non-affine deformations down to the mesh scale.

\smallskip
\noindent
{\it Acknowledgments\/}---%
We thank Tom Lubensky and Swagatam Mukhopadhyay for discussions.
This work was supported by
NSF DMR
02-05858 and
06-05816
(XM, PMG and XX),
ACS PRF (XX),
and
DFG through SFB~602 and Grant No.~Zi~209/7 (AZ).
PMG and XX thank for its hospitality the Aspen Center for Physics.

\end{document}